\documentclass[conference]{IEEEtran}
\IEEEoverridecommandlockouts
\usepackage{cite}
\usepackage{amsmath,amssymb,amsfonts}
\usepackage{algorithmic}
\usepackage{graphicx}
\usepackage{textcomp}
\usepackage{caption}
\usepackage{subcaption}
\usepackage{latexsym}
\usepackage{amsmath}
\usepackage[mathscr]{euscript}
\usepackage[ruled]{algorithm2e}
\SetArgSty{textnormal}
\usepackage{threeparttable}
\usepackage{threeparttablex}
\usepackage{commath}
\usepackage{mathtools}
\usepackage{slashbox}
\usepackage{pict2e}
\usepackage{epstopdf}
\usepackage{verbatim}
\usepackage{subfloat}
\usepackage{tabu}
\usepackage{booktabs}
\usepackage{float}
\usepackage{amssymb}
\usepackage{url}
\usepackage{multirow}
\usepackage{graphicx}
\usepackage{mathptmx}% use Times fonts if available on your TeX system
\usepackage{cite}
\usepackage{footnote}
\usepackage{array}
\usepackage[table]{xcolor}
\usepackage{tabularx}

\usepackage{amsthm} % for proof
\theoremstyle{theorem}
\allowdisplaybreaks
\def\BibTeX{{\rm B\kern-.05em{\sc i\kern-.025em b}\kern-.08em
    T\kern-.1667em\lower.7ex\hbox{E}\kern-.125emX}}

\begin{document}

\title{Energy-Efficient Precoding for Multi-User Visible Light Communication with Confidential Messages}

\author{\IEEEauthorblockN{Son T. Duong\IEEEauthorrefmark{1},
Thanh V. Pham\IEEEauthorrefmark{2},
Chuyen T. Nguyen\IEEEauthorrefmark{1}, and
Anh T. Pham\IEEEauthorrefmark{2}}
\IEEEauthorblockA{\IEEEauthorrefmark{1}Hanoi University of Science and Technology, Vietnam.}
\IEEEauthorblockA{\IEEEauthorrefmark{2}Computer Communications Lab., The University of Aizu, Japan.}
Emails: \IEEEauthorrefmark{1}duongthanhson2808@gmail.com, \IEEEauthorrefmark{1}chuyen.nguyenthanh@hust.edu.vn, \IEEEauthorrefmark{2}\{tvpham, pham\}@u-aizu.ac.jp
% \IEEEauthorblockA{\IEEEauthorrefmark{3}Hanoi University of Science and Technology, Vietnam}
% \IEEEauthorblockA{\IEEEauthorrefmark{4}Computer Communications Lab, University of Aizu, Aizuwakamatsu, Japan}% <-this % stops an unwanted space
}

\maketitle

\begin{abstract}
% This document is a model and instructions for \LaTeX.
% This and the IEEEtran.cls file define the components of your paper [title, text, heads, etc.]. *CRITICAL: Do Not Use Symbols, Special Characters, Footnotes, 
% or Math in Paper Title or Abstract.
In this paper, an energy-efficient precoding scheme is designed for multi-user visible light communication (VLC) systems in the context of physical layer security, where users' messages are kept mutually confidential. The design problem is shown to be non-convex fractional programming, therefore Dinkelbach algorithm and convex-concave procedure (CCCP) based on the first-order Taylor approximation are utilized to tackle the problem. Numerical results are performed to show the convergence behaviors and the performance of the proposed solution for different parameter settings.
\end{abstract}


\section{Introduction}
Visible light communication (VLC) has been becoming an attractive wireless solution to complement existing technologies due to its  high-capacity data transmission with license-free spectrum. Aside from this, the technology also takes advantage of the widespread deployment of light-emitting diodes (LEDs). This naturally enables VLC to fit into the future ubiquitous networks.

%high energy-efficient light-emitting diodes (LEDs), which are widely deployed for indoor illumination. %This naturally enables VLC to fit into the future ubiquitous networks.
%The exponential growth of Internet data traffic over the past decade has been motivating a great deal of research in new wireless technologies. In this regard, visible light communication (VLC) has been recognized one of the most promising solutions in providing high-capacity data transmission with license-free spectrum. Aside from this, the technology also takes advantage of the high energy-efficient light-emitting diodes (LEDs), which are widely deployed for indoor illumination. This naturally enables VLC to fit into the future ubiquitous networks.

However, multiple challenges still remain to make VLC more viable, among which security in terms of information privacy and confidentiality (especially in public areas) are of the most important issues \cite{arfaoui2020}. In this respect, physical layer security (PLS) has emerged as novel paradigm to enhance secure communication  by exploiting the randomness of the wireless channels, noise, and interference. The most promising implication of PLS is that a perfect secured communication can be achieved from information theoretic point of view. The secrecy of PLS is quantized by the secrecy rate that defines the maximum transmission rate at which unauthorized users are unable to extract any information from the received signals regardless of their computational capability.
While PLS has been a well investigated topic in radio frequency (RF) communications, it has only been receiving considerable attention in the past few years in the case of VLC. %Compared to RF links, VLC channels exhibit several unique characteristics in terms of power constraints, which make an exact expression of the secrecy rate unavailable.
% Considering a single-input single-output (SISO) VLC channel, which comprises a transmitter, a legitimate user, and an eavesdropper, lower and upper bounds secrecy rate were derived in \cite{wang2018physical}. In case of one transmitter and multiple receivers including legitimate users and eavesdroppers, non-orthogonal multiple access scheme (NOMA) was investigated for enhancing the secrecy performance of the system in \cite{zhao2018physical}.
In practical VLC systems, multiple LED luminaries should usually be deployed to provide a sufficient illumination. As such, multiple-input single-output (MISO) channels are more prevalent. In such scenarios, the degrees of freedom introduced by multiple LED transmitters enables the use of precoding as a means of secrecy enhancement \cite{mostafa2015physical,mostafa2016optimal,ma2016optimal,pham2017secrecy,arfaoui2017achievable,arfaoui2018secrecy,cho2018,Cho2021}. 

Aside from the security, energy efficiency is another essential criterion in designing a communications system. This increasing attention to energy consumption comes from the current global effort to reduce the carbon footprint. In regard to PLS in VLC, there has been a few studies dealing with precoding design from the energy consumption perspective. In the case of one legitimate user and multiple eavesdroppers, the authors in \cite{ma2016optimal} and \cite{liu2020beamforming} considered the problem of minimizing the power of the information-bearing signal while ensuring a minimum achievable secrecy rate. With the same configuration, \cite{pham2020,pham2020energy} focused on designing artificial noise-aided precoding. The total power consumption of the information-bearing signal and the artificial noise is then minimized taking into account predefined thresholds on the signal-to-interference-plus-noise ratios (SINRs) of the legitimate user and eavesdroppers.    

In general, the above mentioned works concerned with the issue of minimizing the consumed power given that a certain secrecy requirement is fulfilled. This design approach, however, is not necessarily optimal from the perspective of energy efficiency, which is defined as the number of bits that can be transmitted per Joule. In other words, it is the ratio of the achievable (secrecy) rate to the total power consumption. Considering that an energy-efficient precoding design for PLS in multi-user (MU) VLC systems has not been studied, the objective of this paper  is to fill this gap.
Particularly, we examine an MU-MISO VLC system where each user treats others as eavesdroppers, hence its intended message must be kept confidential. The energy efficiency in the context of PLS of such system is then formulated using a lower bound on the achievable secrecy sum-rate. Regarding the power consumption, in addition to the power consumed for the information-bearing signal and circuitry operations (e.g., as in radio frequency (RF) communications), the lighting function of VLC requires an additional power for illumination, which also impacts the achievable secrecy rate. The optimal precoding design problem to maximize the energy efficiency is shown to be non-convex fractional programming, which renders finding an exact solution difficult. Hence, our approach is to find a sub-optimal yet computational efficient solution by employing the Dinkelbach algorithm and convex-concave procedure (CCCP). Simulations under different system parameters are conducted to verify the efficiency of the proposed solution.  

The rest of the paper is structured as follows. The system model together with a formulation of the energy efficiency are described in Section II. Section III focuses on solving the precoding design to maximize the energy efficiency. Numerical results and related discussions are given in Section IV. Finally, Section V concludes the paper. 

Notation: $\mathbb{R}^{m \times n}$ denotes the $m \times n$ real-valued matrices. %The symbol $\mathbb{S}^m$ represents the set of symmetric matrices with the size of $n \times n$. 
%The lowercase symbol $s$ represents scalar variable. 
The uppercase bold symbols, e.g. $\mathbf{M}$, denote matrices, while the lowercase ones such as $\mathbf{v}$ represent column vectors. The transpose of $\mathbf{M}$ is written as $\mathbf{M}^T$. The $k$-th column vector of $\mathbf{M}$ is denoted as $\mathbf{m}_k$, while its ($i,j$)-th element is written as $m_{i,j}$. %A series of matrices $\mathbf{A}$ (or a one-dimensional array of matrices $\mathbf{A}$) can be denoted as $\mathbf{A}^{(1)}$, $\mathbf{A}^{(2)}$, ..., $\mathbf{A}$. In order to extend this array into two-dimensional array of matrices, each element (matrix) of two-dimensional array is denoted as $\mathbf{A}[m]$. 
Additionally,  $|\cdot|_1$, $\lVert\cdot\rVert$ and $|\cdot|$ are ${L}_1$ norm, the Euclidean norm and absolute value operator, respectively. 
%$\mathbf{I}_N$ and $\mathbf{e}_n$ denotes the identity matrix of size $N$ and all-zero vector except the $n$-th element being 1.
\section{System Model}
\label{sec:model}
%=========================================================================================================
%\subsection{VLC System Description}
% \begin{figure*}
% \centering
% \includegraphics[scale=0.45]{linktiming.eps}
% \caption{Link timing between the reader and tags with collision/empty/success slots.}
% \label{fig:linktiming}
% \end{figure*}
\begin{figure}[htbp]
\centerline{\includegraphics[scale = 0.22]{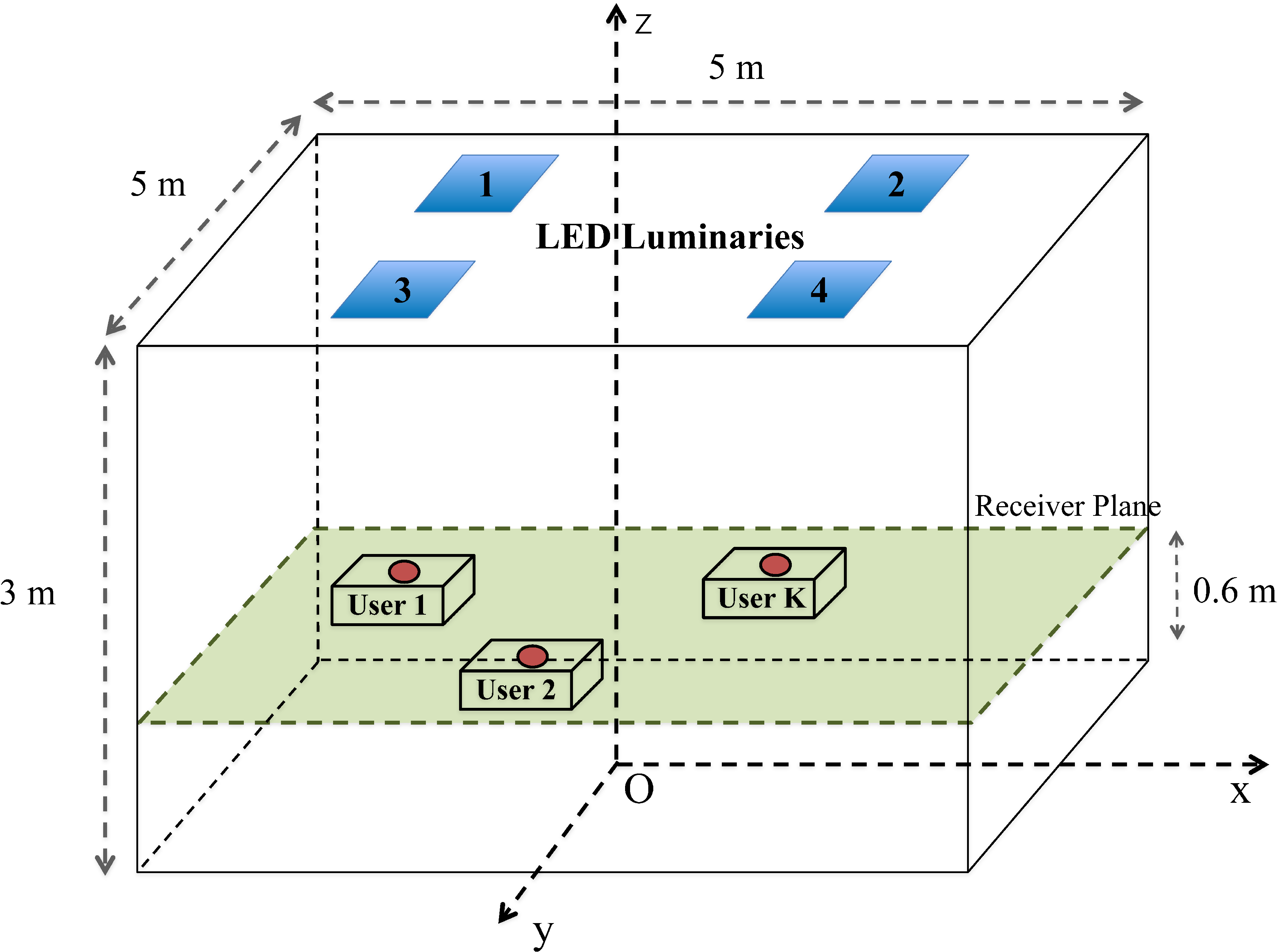}}
\caption{A simple example of the considered MU-MISO VLC system with $N_T=4$ LED arrays and $K=3$ users.}
\label{system_model}
\end{figure}
Our considered MU-MISO VLC system as illustrated in Fig.~\ref{system_model} consists of $N_T$ LED luminaries and $K$ decentralized users, where each user is equipped with a photodiode (PD). It is reminded that the transmission is considered to be confidential if each user is not able to decode any information intended to the others. 
%----------------------------------------------------------------------------------------------
\subsection{Signal Model}
Let $\mathbf{d} = \begin{bmatrix}d_1 & d_2 & \cdots & d_K\end{bmatrix}^T \in \mathbb{R}^{K \times 1}$ be the vector of data symbols for all users. Assume that the symbols are drawn from an $M$-ary pulse amplitude modulation ($M$-PAM) and are modeled as a random variable (RV) $d$ following a certain distribution over $[-1, 1]$ with zero-mean and variance $\sigma^2_d$. 
%For the broadcast transmission, At the output of driver at the $n$-th LED array, 
An information-bearing signal $s_n$ for the $n$-th LED transmitter is generated from a linear combination of the data vector and a precoder $\mathbf{v}_n = \begin{bmatrix}w_{n,1} & w_{n,2} & \cdots & w_{n,K}\end{bmatrix} \in \mathbb{R}^{1 \times K}$ as
\begin{equation}
\label{eqn:information_bearing_signal}
    s_n = \mathbf{v}_n \ \mathbf{d}.
\end{equation}
For illumination, an DC bias $I_{n}^{\text{DC}}$ should be added to $s_n$ to create a non-negative drive current $x_n$ for the LED. The drive current, in addition, needs to be constrained to a maximum threshold, i.e., $I_{\text{max}}$, to ensure that LEDs operate normally. Therefore,   
%Since $d_i$ is in the range of [-1,1], the broadcast signal $s_n$ might be negative. In order to satisfy a non-negative constraint on LED's drive current, a DC's bias should be added as follows
\begin{equation}
%\label{eqn:transmitted_signal}
\label{eqn:Imax}
    0 \leq x_n=s_n+I_n^{\text{DC}} \leq I_{\text{max}}.
\end{equation}
%where $I_n^{\text{DC}}$ is the DC-bias for the $n$-th LED array, which, along with $\mathbf{v}_n$, is chosen to ensure $u_n \geq 0$. The drive current should be also smaller than a threshold denoted by $I_{\text{max}}$ to to guarantee a normal operation of the LEDs, i.e., avoid the overheating of the LEDs and the potential light intensity reduction, as follows 
%\begin{equation}
%\label{eqn:Imax}
%    0 \leq s_n + I_n^{\text{DC}} \leq I_{\text{max}}.
%\end{equation}
%Moreover, let $\mathbf{P}^s=[P_1^s \ P_2^s \ ... \ P_{N_T}^s]$ define LED arrays' emitted optical power vector, where $P_n^s$ is for the $n$-th LED array. Then, $P_n^s=\eta u_n$ where $\eta$ is the LED conversion factor. On the other hand, since $\mathbb{E}[s_n]=0$, $I_n^{\text{DC}}$ uniquely determines the average illuminance power of the $n$-th LED array denoted by $\overline{P_n^s}$ as follows
The emitted optical power of each LED luminary is given by
\begin{equation}
    {P}_n^s=\eta \left(s_n+I_n^{\text{DC}}\right),
\end{equation}
where $\eta$ is the LED conversion factor. %, calculated by the ratio of illuminance energy emitted by LED and drive current.
% \begin{figure}[htbp]
% \centerline{\includegraphics[scale=0.65]{MU_MISO_VLC_2.jpg}}
% \caption{Signal transmission schematic diagram with precoding.}
% \label{MU_MISO_VLC_diagram}
% \end{figure}
Denote $\mathbf{h}_k = [h_{1,k} \ h_{2,k} \ \cdots \ h_{N_T,k}]^T$ is the $k$-th user's channel matrix, where $h_{n, k}$ is the line-of-sight (LoS) channel coefficient between the $n$-th LED array and the user. Details on the VLC channel model can be found in \cite{arfaoui2020} and references therein. 
% The received power at the $k$-th user is represented as
% \begin{equation}
% \label{eqn:received_power}
%     P_k^r=\mathbf{h}_k^T \ \mathbf{P}^s,
% \end{equation}
% where $\mathbf{h}_k = [h_{1,k} \ h_{2,k} \ ... \ h_{N_T,k}]^T$ is the $k$-th user's channel matrix. Then, if $\gamma$ is denoted as the PD responsivity, 
The electrical signal at the PD output is then given by
\begin{align}
\label{eqn:received_signal}
    y_k &=  \gamma \mathbf{h}^T_k\begin{bmatrix}P_1^s & P_2^s & \cdots & P^s_{N_T,k}\end{bmatrix}  + n_k \nonumber \\
        %= & \gamma H_k^T \eta u_n + n_k \\
        &= \gamma \eta \left(\mathbf{h}_k^T \mathbf{w}_k d_k + \mathbf{h}_k\sum_{i=1,\ i\neq k}^{K} \mathbf{w}_i d_i + \mathbf{h}_k \mathbf{I}^{\text{DC}}\right) + n_k,
\end{align}
%where $u=[u_1 \ u_2 \ ... \ u_{N_T}]^T$ and 
where $\mathbf{w_k} = \begin{bmatrix}w_{1,k} \ w_{2, k} \ \cdots \ w_{N_T, k}\end{bmatrix}^T$ is the $k-$th user's precoder and $\mathbf{I}^{\text{DC}} = \begin{bmatrix}I_1^{\text{DC}} \ I_2^{\text{DC}} \ \cdots \ I_{N_T}^{\text{DC}}\end{bmatrix}^T$. It is noted that since $|d_i| \leq 1$, we have 
\begin{align}
    \label{eqn:cond2}
    -\lVert\mathbf{v}_k\rVert_1\leq s_n\leq \lVert\mathbf{v}_k\rVert_1.
\end{align}
To ensure both (\ref{eqn:Imax}) and (\ref{eqn:cond2}), the following constraint should be satisfied
\begin{equation}
    \sum_{k=1}^K |w_{n,k}| \leq \min\left(I_n^{\text{DC}},I_{\text{max}}-I_n^{\text{DC}}\right).
    %= \min\left(I_n^{\text{DC}},I_{\text{max}}-I_n^{\text{DC}}\right),
\end{equation}
%where $\min\left(I_n^{\text{DC}},I_{\text{max}}-I_n^{\text{DC}}\right) \triangleq \min(I_n^{\text{DC}},I_{\text{max}}-I_n^{\text{DC}})$. 
%are transmitted signal vector and the DC-bias current vector, respectively. 

The receiver noise $n_k$ in \eqref{eqn:received_signal} can be modeled as a real-valued zero-mean Gaussian RV whose variance is given by
\begin{equation}
\label{eqn:noise_var}
    \sigma_k^2 = 2\gamma e \overline{P}_k^r B + 4\pi e A_r \gamma \chi_{\text{amb}}(1-\cos(\Psi))B + i_{\text{amp}}^2B,
\end{equation}
where $\overline{P}_k^r = \eta\mathbf{h}_k^T\mathbf{I}^{\text{DC}}$ is the average of received power at the $k$-th user, $A_r$ is the area of the PD, $e$ is the elementary charge, $\Psi$ is the optical field of view (FoV) ofthe PD, $B$ is the modulation bandwidth, $\chi_{\text{amb}}$ is the ambient light photo-current, and $i_{\text{amp}}$ is the pre-amplifier noise current density. 

%Besides, $\mathbf{w}_k$ is the precoder for the $k$-th user and is given by $\mathbf{w}_k = [w_{1,k} \ w_{2,k} \ ... \ w_{N_T,k}]^T$, while the precoding matrix $\mathbf{W}=[\mathbf{w}_1 \ \mathbf{w}_2 \ ... \ \mathbf{w}_K]$ = $[\mathbf{v}_1^T \ \mathbf{v}_2^T \ ... \ \mathbf{v}_{N_T}^T]^T$. 

\subsection{Power Consumption}
We now analyze the total consumed power at LED transmitters, which can be expressed by %On the other hand, the total power consumption at the transmitter side is determined as follows
\begin{equation}
\label{eqn:power_consumption}
    P_{\text{total}} = P_{\text{DC}}+ P_{\text{AC}},
\end{equation}
where $P_{\text{DC}}$ and $P_{\text{AC}}$ are the powers of the DC and AC currents, respectively.
The DC current power includes the power used for illumination by LEDs denoted as  $P_{\text{DC, LEDs}}$ and that used by other circuit components denoted as $P_{\text{DC, circuitry}}$. While  $P_{\text{DC, circuitry}}$ can be considered to be fixed, the power consumption of LEDs can be adjusted depending on the required dimming level. However, under a specific usage when the illumination level is not changed, it is reasonable to assume that $P_{\text{DC, LEDs}}$ is fixed as well. 
%Of course, in specific usage when the LED's optical power are not changed, we can assume that the total DC power consumption of the system is fixed. Let's the DC power consumption of LED being $P_{\text{DC,led}}$ and the DC power consumption of remaining components being $P_{\text{DC,fixed}}$. 
The DC power consumption is then given as
\begin{equation}
\label{eqn:DC_consumption}
    P_{\text{DC}} = P_{\text{DC, LEDs}}+ P_{\text{DC,circuitry}},
\end{equation}
with $P_{\text{DC, LEDs}}$ being calculated as
\begin{equation}
\label{eqn:DC_led_consumption}
    P_{\text{DC, LEDs}} = \sum_{n=1}^{N_T} \ U_{\text{LED}} \ I_n^{\text{DC}},
\end{equation}
where $U_{\text{LED}}$ being the forward voltage of the LEDs.

The AC currents comes from the output current (or voltage) from the precoders of LED drivers, therefore can be calculated as
\begin{equation}
\label{eqn:LED_drive_power_consumption}
    P_{\text{AC}} = r \sum_{k=1}^K \sigma_d^2 \lVert\mathbf{w}_k\rVert^2,
\end{equation}
where $r$ is the equivalent resistance of the AC circuit. %components where AC current flows through. %, $\sigma_d^2$ is the variance of data symbol $d$. Because symbol $d$ is uniformly distributed between [-1,1], $\sigma_d^2=1/3$. 
Without loss of generality, we denote $\xi = r \sigma_d^2$ as equivalent resistance, then rewrite the total power consumption as
\begin{equation}
\label{eqn:power_consumption2}
    P_{\text{total}} = P_{\text{DC}} + \xi \sum_{k=1}^K \lVert\mathbf{w}_k\rVert^2.
\end{equation}
\subsection{Energy Efficiency}
For demodulation, the DC component in the received signal in \eqref{eqn:received_signal} is removed, yielding
\begin{align}
\overline{y}_k = \mathbf{h}_k^T \mathbf{w}_k d_k + \mathbf{h}_k\sum_{i=1,\ i\neq k}^{K} \mathbf{w}_i d_i  + \overline{n}_k,
\end{align}
where $\overline{n}_k = \frac{n_k}{\gamma \eta}$.
According to \cite{arfaoui2018secrecy}, a lower bound of achievable confidential secrecy rate of the $k-$th user is given as
\begin{align}
\label{eqn:secrecy_rate}
	\begin{split}
        R_{s,k}(\mathbf{W}) = & \frac{1}{2}\log_2\left(\frac{1+\sum_{i=1}^K a_k\left(\mathbf{h}_k^T \mathbf{w}_i\right)^2}{1+\sum_{i=1,i\neq k}^K b_k\left(\mathbf{h}_k^T \mathbf{w}_i\right)^2}\right) \nonumber \\ &  - \frac{1}{2}\log_2\left(1+\sum_{i=1,i\neq k}^K b_i\left(\mathbf{h}_i^T \mathbf{w}_k\right)^2\right),
     	\end{split}
\end{align}
where $\mathbf{W}=\begin{bmatrix}\mathbf{w}_1 \ \mathbf{w}_2 \ \cdots \ \mathbf{w}_K\end{bmatrix}$, 
$a_k=\frac{\exp(2h_d)}{2\pi e \overline{\sigma}_k^2}$, and $b_k=\frac{\sigma_d^2}{\overline{\sigma}_k^2}$ with $\overline{\sigma}_k^2 = \frac{\sigma^2_k}{(\gamma\eta)^2}$ and $h_d$ being the differential entropy of $d$.
%and the variance of the random scalar variable $u$, respectively. $\sigma_i^2$ denotes by the noise variance of the $i$-th user. 
The energy efficiency with respect to the achievable secrecy sum-rate of the considered system is therefore given by
\begin{equation}
\label{eqn:energy_efficiency}
    \Phi(\mathbf{W})=\frac{\sum_{k=1}^K R_{s,k}(\mathbf{W})}{P_{\text{DC}} + \xi \text{Tr}\left(\mathbf{W}\mathbf{W}^T\right)}.
\end{equation}
%----------------------------------
\section{Sub-optimal Precoding Design}
Our design objective is to maximize the energy efficiency while a minimum achievable secrecy rate for each user is guaranteed. Hence, the design problem can be formulated as follows 
\begin{subequations}
\label{eqn:optimization_problem_ini}
    \begin{align}
        \max_{\mathbf{W}} \ \ & \ \Phi(\mathbf{W}), \label{eqn:optimization_problem.a}\\
        \text{s.t.} \ \ & \ R_{s,k}(\mathbf{W})  \geq \lambda_k,  \label{eqn:gamma_k}\\
            & \sum_{k=1}^K |w_{n,k}|  \leq \min\left(I_n^{\text{DC}},I_{\text{max}}-I_n^{\text{DC}}\right) \label{eqn:linear_constraints} ,
    \end{align}
\end{subequations}
where $\lambda_k$ is the threshold for the secrecy rate of the $k$-th user.
It is seen from (\ref{eqn:energy_efficiency}) that the objective function is a non-concave fractional function with $\mathbf{W}$. This  requires the use of Dinkelbach algorithm \cite{dinkelbach1967nonlinear}, which is efficient in solving fractional programming. Let $N(\mathbf{W})=\sum_{k=1}^K R_{s,k}(\mathbf{W})$, $D(\mathbf{W}) = P_{\text{DC}} + \xi \text{Tr}\left(\mathbf{W}\mathbf{W}^T\right)$. Then for some $\mu \geq 0$, the Dinkelbach algorithm involves solving the following  
\begin{subequations}
\label{eqn:Dinkelbach_problem}
    \begin{align}
        \max_{\mathbf{W}} \ \ & N\left(\mathbf{W}\right) - \mu D\left(\mathbf{W}\right) ,\\
        \text{s.t.} \ \ & \ R_{s,k}(\mathbf{W})  \geq \lambda_k,  \label{eqn:dink_gamma_k}\\
            & \sum_{k=1}^K |w_{n,k}|  \leq \min\left(I_n^{\text{DC}},I_{\text{max}}-I_n^{\text{DC}}\right) . %\label{eqn:linear_constraints} .
    \end{align}
\end{subequations}
to obtain a precoder $\mathbf{W}'$. The optimal value $\mu^{*} = \frac{N(\mathbf{W^*})}{D(\mathbf{W^*})}$ is achieved when  $N(\mathbf{W'}) - \mu D(\mathbf{W'}) = 0$. This results in the Dinkelbach algorithm described as follows

\begin{algorithm}[h]
\SetAlgoLined % activate/deactivate number line
%\KwData{this text}
%\KwResult{how to write algorithm with \LaTeX2e }
%\algrule[0.5pt]
Choose the maximum number of iterations ${{L}}_{\text{max}, 1}$ and the error tolerance $\epsilon_1$.\\
Initialize $\mu>0$, $l=0$.\\
\While{$\text{convergence}==\bold{False} \ \text{and} \ l \leq {L}_{\text{max}, 1}$}{
    For a given $\mu$, solve (\ref{eqn:Dinkelbach_problem}) to get  $\mathbf{W}^{(l)}$.\\
    \eIf{ $N\left(\mathbf{W}^{(l)}\right)-\mu D\left(\mathbf{W}^{(l)}\right) \leq \epsilon_1$ }{
        $\text{convergence}==\mathbf{True}$;\\
        ${\mathbf{W}^*}={\mathbf{W}^{(l)}}$; \\
        $\mu^*=\frac{N\left(\mathbf{W}^{(l)}\right)}{D\left(\mathbf{W}^{(l)}\right)}$;
    }
    {
        $\text{convergence}==\mathbf{False}$ \\
        $l=l+1$; \\
        Update $\mu = \frac{N\left(\mathbf{W}^{(l)}\right)}{D\left(\mathbf{W}^{(l)}\right)}$;
    }
}
Return the optimal ${\mathbf{W}^*}$ and  ${\mu^*}$.
\caption{Dinkelbach-type algorithm for solving \eqref{eqn:Dinkelbach_problem}.}
\label{Alg1}
\end{algorithm}
Nevertheless, it should be noted that (\ref{eqn:Dinkelbach_problem}) is not a convex optimization problem due to the non-concave objective function and the non-convex constraint in \eqref{eqn:dink_gamma_k}. To overcome this problem, we make use of the CCCP approach based on the first-order Taylor approximation to approximate the original problem to a convex one. Specifically, we first introduce the following slack variables %$\mathbf{R}=\begin{bmatrix}\mathbf{r}_1\ \mathbf{r}_2 \ \mathbf{r}_3\end{bmatrix} \in \mathbb{R}^{3\times K}$  and $\mathbf{P}=\begin{bmatrix}\mathbf{p}_1 \ \mathbf{p}_2 \ \mathbf{p}_3\end{bmatrix} \in \mathbb{R}^{3\times K}$, each of those vectors consists of elements considered as slack variables and given as follows
\begin{subequations}
    \begin{align}
        r_{1,k} \ \overset{\Delta}{=} \ & \frac{1}{2}\log_2\left(1+\sum_{i=1}^K a_k\left(\mathbf{h}_k^T \mathbf{w}_i\right)^2\right) \label{eqn:r1k} ,\\
        p_{1,k} \ \overset{\Delta}{=} \ & \sum_{i=1}^K a_k\left(\mathbf{h}_k^T \mathbf{w}_i\right)^2 \label{eqn:p1k} ,\\
        r_{2,k} \ \overset{\Delta}{=} \ & \frac{1}{2}\log_2\left(1+\sum_{i=1,i\neq k}^K b_k\left(\mathbf{h}_k^T \mathbf{w}_i\right)^2\right) \label{eqn:r2k} ,\\
        p_{2,k} \ \overset{\Delta}{=} \ & \sum_{i=1,i\neq k}^K b_k\left(\mathbf{h}_k^T \mathbf{w}_i\right)^2 \label{eqn:p2k} ,\\
        r_{3,k} \ \overset{\Delta}{=} \ & \frac{1}{2}\log_2\left(1+\sum_{i=1,i\neq k}^K b_i\left(\mathbf{h}_i^T \mathbf{w}_k\right)^2\right) \label{eqn:r3k} ,\\
        p_{3,k} \ \overset{\Delta}{=} \ & \sum_{i=1,i\neq k}^K b_i\left(\mathbf{h}_i^T \mathbf{w}_k\right)^2 \label{eqn:p3k},
    \end{align}
\end{subequations}
Then, the objective function $\sum_{k=1}^K\left(r_{1,k}-r_{2,k}-r_{3,k}\right)-\mu\left(P_{\text{DC}}+\xi \text{Tr}\left(\mathbf{W}\mathbf{W}^T\right)\right)$ is a concave function with respect to $\mathbf{W}$, $r_{1,k}$, $r_{2,k}$, and $r_{3,k}$. Also, (\ref{eqn:Dinkelbach_problem}) can be rewritten as 
\begin{subequations}
\label{eqn:Dinkelbach_problem_2}
    \begin{align}        \max_{\substack{\mathbf{W},r_{1, k}, r_{2, k}, r_{3, k} \\  p_{1, k}, p_{2, k}, p_{3, k}}} \ & \sum_{k=1}^K\left(r_{1,k}-r_{2,k}-r_{3,k}\right)-\mu\left(P_{\text{DC}}+\xi \text{Tr}\left(\mathbf{W}\mathbf{W}^T\right)\right) , \label{eqn:concave_objective_function}\\
        \text{s.t.} \ \ & r_{1,k} \leq \frac{1}{2}\log_2\left(1+p_{1,k}\right) ,            \label{eqn:r_1k_constraints} \\
                & p_{1,k} \leq \sum_{i=1}^K a_k\left(\mathbf{h}_k^T \mathbf{w}_i\right)^2 ,                 \label{eqn:p_1k_constraints} \\
                & r_{2,k} \geq \frac{1}{2}\log_2\left(1+p_{2,k}\right) ,            \label{eqn:r_2k_constraints} \\
                & p_{2,k} \geq \sum_{i=1,i\neq k}^K b_k\left(\mathbf{h}_k^T \mathbf{w}_i\right)^2 ,         \label{eqn:p_2k_constraints}\\
                & r_{3,k} \geq \frac{1}{2}\log_2\left(1+p_{3,k}\right) ,            \label{eqn:r_3k_constraints}\\
                & p_{3,k} \geq \sum_{i=1,i\neq k}^K b_i\left(\mathbf{h}_i^T \mathbf{w}_k\right)^2 ,         \label{eqn:p_3k_constraints}\\
                & r_{1,k}-r_{2,k}-r_{3,k} \geq \lambda_k ,                 \label{eqn:r_constraints}\\
                & \sum_{k=1}^K |w_{n,k}|  \leq \min\left(I_n^{\text{DC}},I_{\text{max}}-I_n^{\text{DC}}\right). \label{eqn:linear_constraints}
    \end{align}
\end{subequations}
It is seen that constraints (\ref{eqn:p_1k_constraints}), (\ref{eqn:r_2k_constraints}), and  (\ref{eqn:r_3k_constraints}) are still not convex. To cope with this, the first-order Taylor approximation is employed to approximate these non-convex constraints. The CCCP is then used to iteratively solve a sequence of approximating convex problems until a predefined convergence criterion is met \cite{yuille2003}. Specifically, at the $m$-th iteration, the following problem is solved
\begin{subequations}
\label{eqn:Dinkelbach_problem_3}
    \begin{align}
       & \max_{\substack{\mathbf{W}, r_{1, k}, r_{2, k}, r_{3, k} \\ p_{1, k}, p_{2, k}, p_{3, k}}} \  \sum_{k=1}^K\left(r_{1,k}-r_{2,k}-r_{3,k}\right)-\mu\left(P_{\text{DC}}+\xi \text{Tr}\left(\mathbf{W}\mathbf{W}^T\right)\right) , \label{eqn:concave_objective_function}\\
       &  \text{s.t.} \nonumber \\ %& r_{1,k} \leq \frac{1}{2}\log[1+p_{1,k}] ,            \label{eqn:r_1k_constraints} \\
                & p_{1,k} \leq \sum_{i=1}^K a_k\left(\left(\mathbf{h}_k^T {\mathbf{w}}_i^{(m-1)}\right)^2 \right. \nonumber \\ & ~~~~~~~\left.+ 2\left({\mathbf{w}}_i^{(m-1)}\right)^T \mathbf{h}_k \mathbf{h}_k^T \left(\mathbf{w}_i^{(m)}-{\mathbf{w}}_i^{(m-1)}\right)\right) , \label{eqn:p_1k_constraints_convex}\\
                & r_{2,k} \geq \frac{1}{2}\log_2\left(1+{p}_{2,k}^{(m-1)}\right) + \frac{\left(p_{2,k}^{(m)}-{p}_{2,k}^{(m-1)}\right)}{2\ln2\left(1+{p}_{2,k}^{(m-1)}\right)} , \label{eqn:r_2k_constraints_convex} \\
                %& p_{2,k} \geq \sum_{i=1,i\neq k}^K b_k(\mathbf{h}_k^T \mathbf{w}_i)^2 ,         %\label{eqn:p_2k_constraints}\\
                & r_{3,k} \geq \frac{1}{2}\log_2\left(1+{p}_{3,k}^{(m-1)}\right) + \frac{\left(p_{3,k}^{(m)}-{p}_{3,k}^{(m-1)}\right)}{2\ln2\left(1+{p}_{3,k}^{(m-1)}\right)} , \label{eqn:r_3k_constraints_con   }\\
                & \eqref{eqn:r_1k_constraints}, \eqref{eqn:p_2k_constraints},\eqref{eqn:p_3k_constraints}-\eqref{eqn:linear_constraints} \nonumber.
                % & p_{3,k} \geq \sum_{i=1,i\neq k}^K b_i(\mathbf{h}_i^T \mathbf{w}_k)^2 ,         \label{eqn:p_3k_constraints}\\
                % & r_{1,k}-r_{2,k}-r_{3,k} \geq \gamma_k ,                 \label{eqn:r_constraints}\\
                % & \sum_{k=1}^K |w_{n,k}|  \leq \min\left(I_n^{\text{DC}},I_{\text{max}}-I_n^{\text{DC}}\right) ,                   \label{eqn:linear_constraints}
    \end{align}
\end{subequations}
where ${\mathbf{w}}_i^{(m-1)}$, ${p}_{2,k}^{(m-1)}$ and ${p}_{3,k}^{(m-1)}$ are the solutions obtained from the previous iteration. Problem (\ref{eqn:Dinkelbach_problem_3}) is convex, thus can be solved efficiently by using standard optimization softwares \textcolor{black}{, such as CVX \cite{cvx}}. Finally, (\ref{eqn:Dinkelbach_problem_2}) can be solved by the proposed {\textbf{Algorithm~\ref{alg.2}}} using CCCP. 
\begin{algorithm}[h]
\SetAlgoLined % activate/deactivate number line
%\KwData{this text}
%\KwResult{how to write algorithm with \LaTeX2e }
%\algrule[0.5pt]
Choose the maximum number of iteration ${L}_{\text{max}, 2}$ and the error tolerance $\epsilon_2>0$. \\
Choose feasible initial points $\mathbf{W}^{(0)}$, $p_{2,k}^{(0)}$, $p_{3,k}^{(0)}$ to (\ref{eqn:Dinkelbach_problem_3}). \\
Set $m=0$. \\
\While{$\text{convergence}==\mathbf{False}$ and $m \leq {L}_{\text{max}, 2}$}{
    $m:=m+1$ \\ 
    Solve  (\ref{eqn:Dinkelbach_problem_3}) for  $\mathbf{W}^{(m)}$, $p_{2,k}^{(m)}$, $p_{3,k}^{(m)}$ using  $\mathbf{W}^{(m-1)}$, $p_{2,k}^{(m-1)}$, $p_{3,k}^{(m-1)}$ obtained from the previous iteration. \\
    \eIf{$\frac{\norm{\mathbf{W}^{(m)} - \mathbf{W}^{(m-1)}}}{\norm{\mathbf{W}^{(m)}}} \leq \epsilon_2$}{
        $\text{convergence} = \mathbf{True}$ \\
        $\mathbf{W}^* = \mathbf{W}^{(m)}$ \\
    }
    {
        $\text{convergence} = \mathbf{False}$. \\
    }
}
Return the optimal value $\mathbf{W}^*$.
\caption{CCCP-type algorithm for solving \eqref{eqn:Dinkelbach_problem_3}}
\label{alg.2}
\end{algorithm}

\section{Simulation Results and Discussions}
\label{sec:simu}
In this section, the performance convergence behaviors of the proposed solution are evaluated. Numerical results are obtained through averaging $10,000$ different users' channel realizations. Unless otherwise noted, the following parameters are used for simulations.  LED bandwidth $B =  20$ MHz,  beam angle $\phi = 120^\circ$,  LED conversion factor $\eta = 2$ W/A, active area of the PD $A_r = 1 \text{cm}^2$,  responsivity $\gamma =  0.54$ A/W, FoV $\Psi =  60^\circ$, optical filter gain $T_s(\psi) =  1$, refractive index of the concentrator $\kappa = 1.5$, ambient light photocurrent $\chi_{\text{amp}} =  10.93 \text{A}/(\text{m}^2 \cdot \text{Sr}$), and  preamplifier noise current density, $i_{\text{amp}} =  5 \text{pA}/\text{Hz}^{-1/2}$. Moreover, the data symbols are assumed to be uniformly distributed over $[-1, 1]$.
%with the proposed precoding designs under different system parameter settings. %The results are obtained by Monte-Carlo method with Matlab simulation.

\begin{figure}[htbp]
    \centering
    \begin{subfigure}[b]{0.3\columnwidth}
        \includegraphics[width=\linewidth]{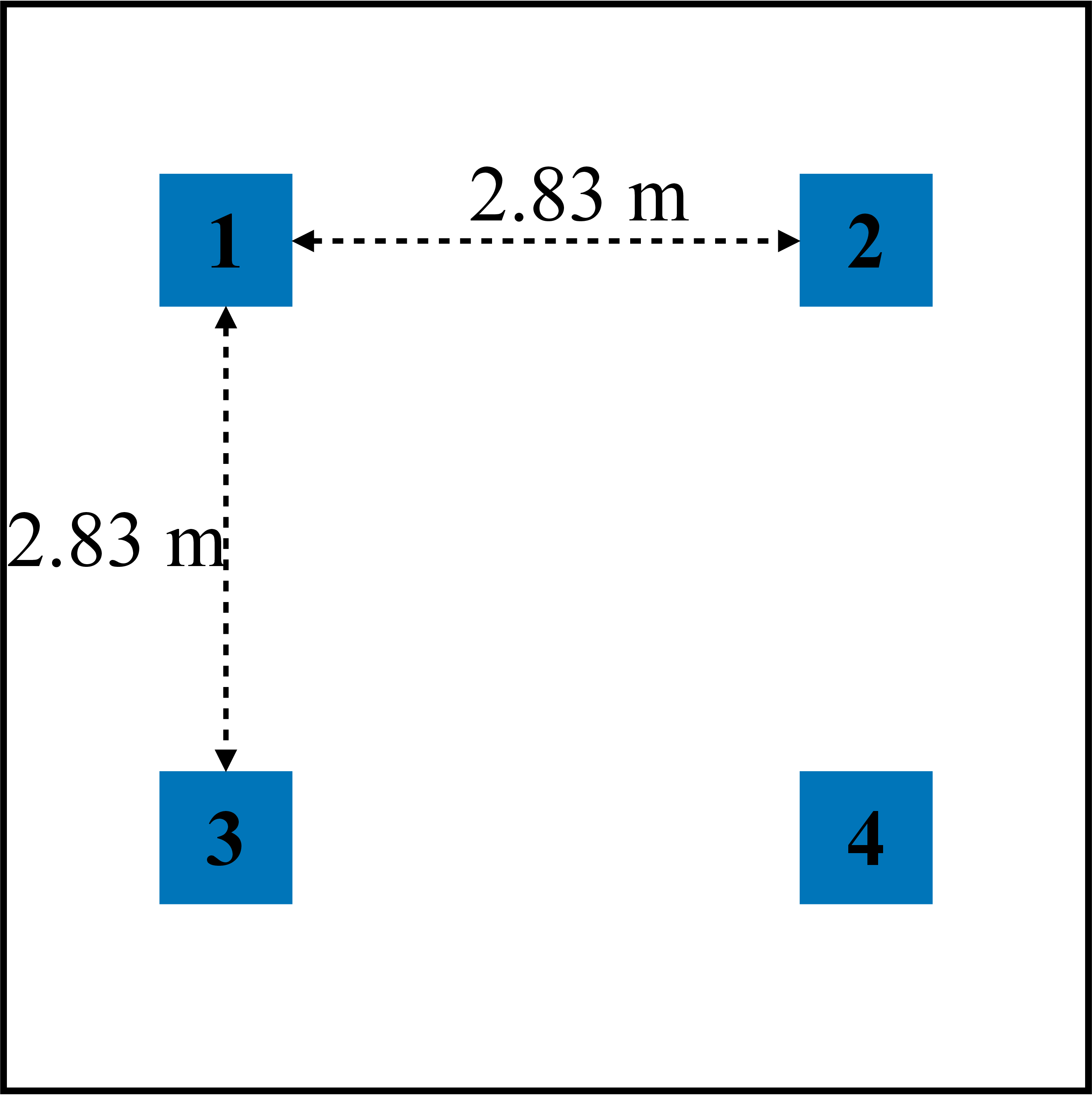}
        \caption{$2\times2$ layout.}
        \label{22}
    \end{subfigure}
    \begin{subfigure}[b]{0.3\columnwidth}
        \includegraphics[width=\linewidth]{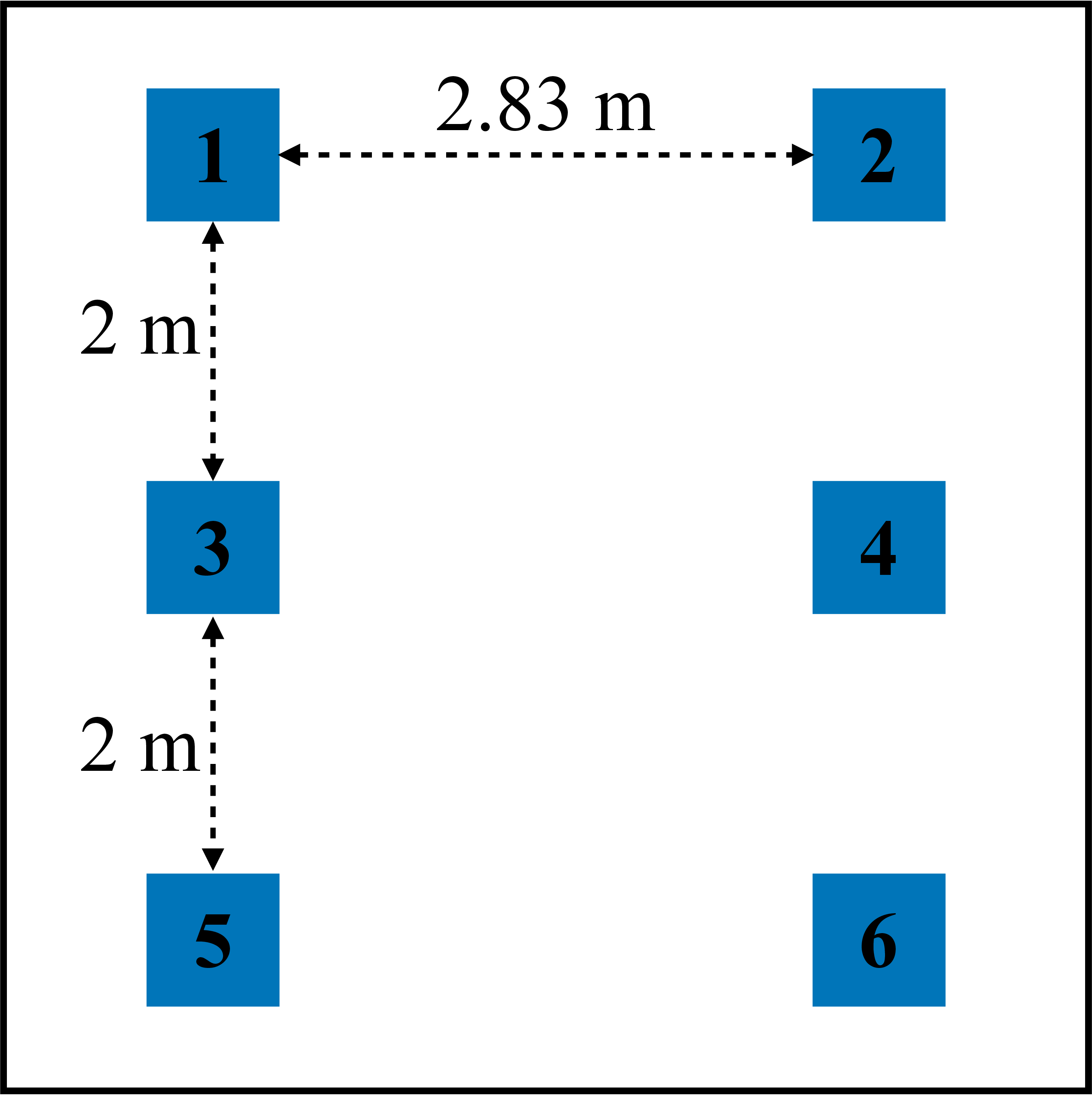}
        \caption{$2\times3$ layout.}
        \label{23}
    \end{subfigure}
    \begin{subfigure}[b]{0.3\columnwidth}
        \includegraphics[width=\linewidth]{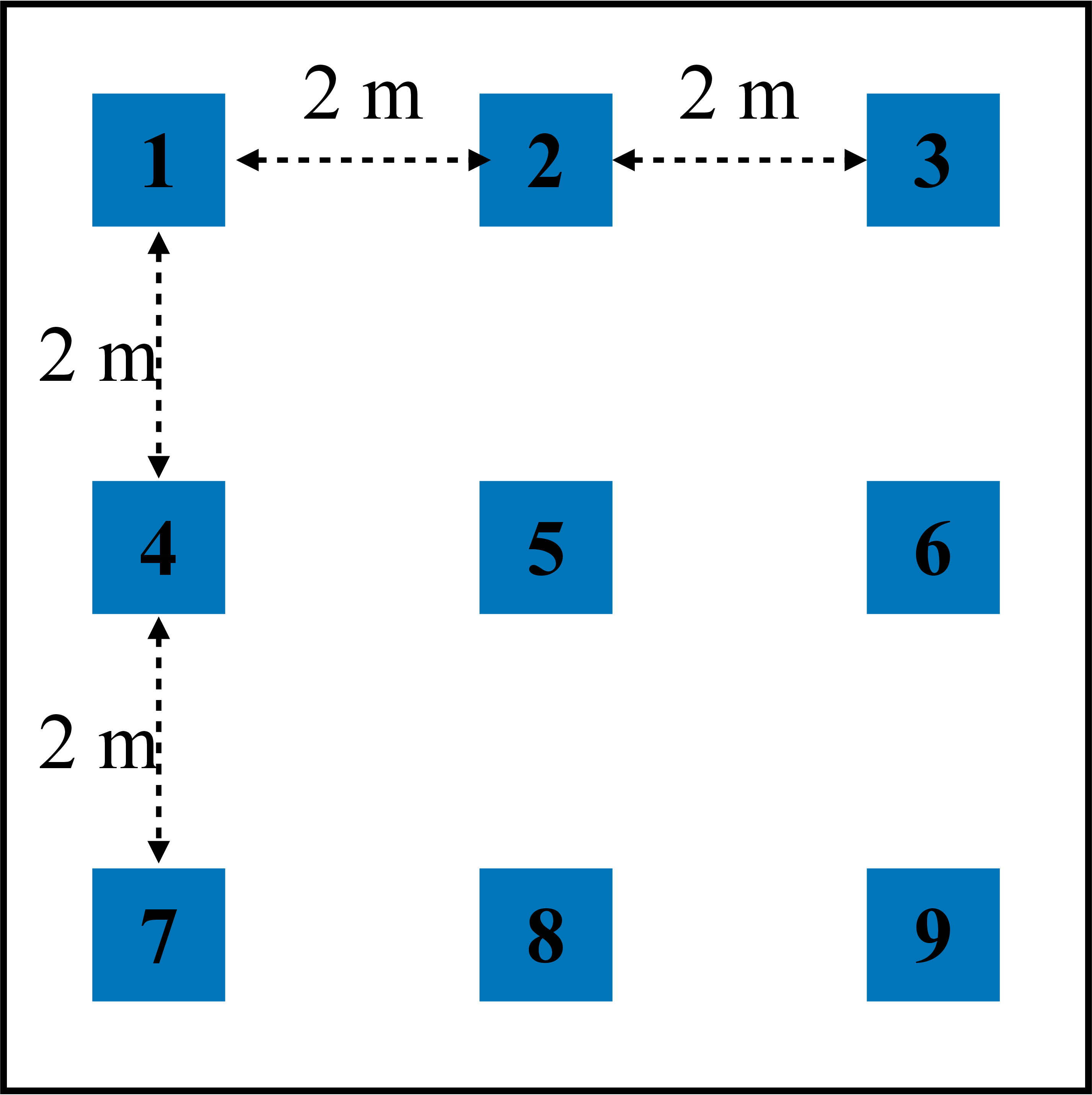}
        \caption{$3\times3$ layout.}
        \label{33}
    \end{subfigure}
    \caption{Different layouts of LED transmitters.}
    \label{LEDlayout}
\end{figure}

\begin{figure}[htbp]
\centerline{\includegraphics[height = 6.0cm, width=8.5cm]{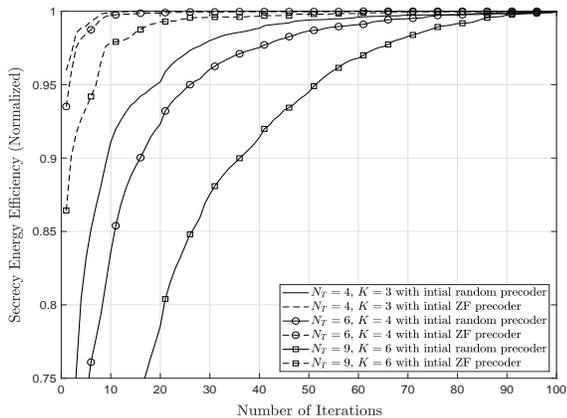}}
\caption{Convergence behaviors of the proposed solution.}
\label{EE_vs_iteration}
\end{figure}

Figure \ref{EE_vs_iteration} illustrates the convergence of the proposed solution. For this purpose, three different scenarios of the numbers of LED transmitters and users, namely $(N_T, K) = (4, 3)$, $(6, 4)$ and $(9, 6)$, are taken into consideration. The configurations of LED transmitter are illustrated in Fig. \ref{LEDlayout}. Additionally, the average emitted optical power of each LED luminary $\overline{P}_n^{s}=\eta I_n^{\text{DC}}=30~\text{dBm}$, $U_{\text{LED}} = 3.3 $V, $P_{\text{DC,circuitry}}=8 \text{W}$, $\xi=3\Omega$, and $\lambda_k$'s $=0.5$ bits/s/Hz are set. 
%In this simulation, we set up the maximum iteration $\text{L}_{\text{max}}=100$, and all algorithms will run multiple times until the number of total iterations reaches 100. Because the size of the system have a big impact on the convergence of our algorithms, we examine three scenarios: 4 transmitters and 3 users, 6 transmitters and 4 users, 9 transmitters and 6 users, whose configurations are shown in the figure (\ref{EE_vs_iteration}), respectively. All other parameters are initialized as defaults. Three scenarios were chosen to guarantee the feasibility of all algorithms being above 90 percents. 
Firstly, we examine the case that the initial precoder $\mathbf{W}^{(0)}$ in {\textbf{Algorithm~\ref{alg.2}}} is chosen randomly. It is shown that considerably large numbers of iterations are required for the normalized energy efficiency to converge (i.e., $60$, $75$, and $90$ iterations  for $(N_T, K) = (4, 3)$, $(6, 4)$ and $(9, 6)$, respectively).  
As we observe that  the optimal solution is usually a near zero-forcing (ZF) precoder, one can choose the initial point as the ZF precoder to speed up the convergence. Indeed, simulation results revealed significant improvements in terms of the required number of iterations. Specifically, using ZF precoder as the initial point, roughly 8, 9, and 20 iterations are needed when $(N_T, K) = (4, 3)$, $(6, 4)$ and $(9, 6)$, respectively. 
%Our results show that the algorithm becomes slower to converge in accordance with the increasing number of transmitters and users. However, if we choose ZF precoder as an initial point to our algorithm, the speed of convergence improves considerably. 

\begin{figure}[htbp]
\centerline{\includegraphics[height = 6.0cm ,width=8.5cm]{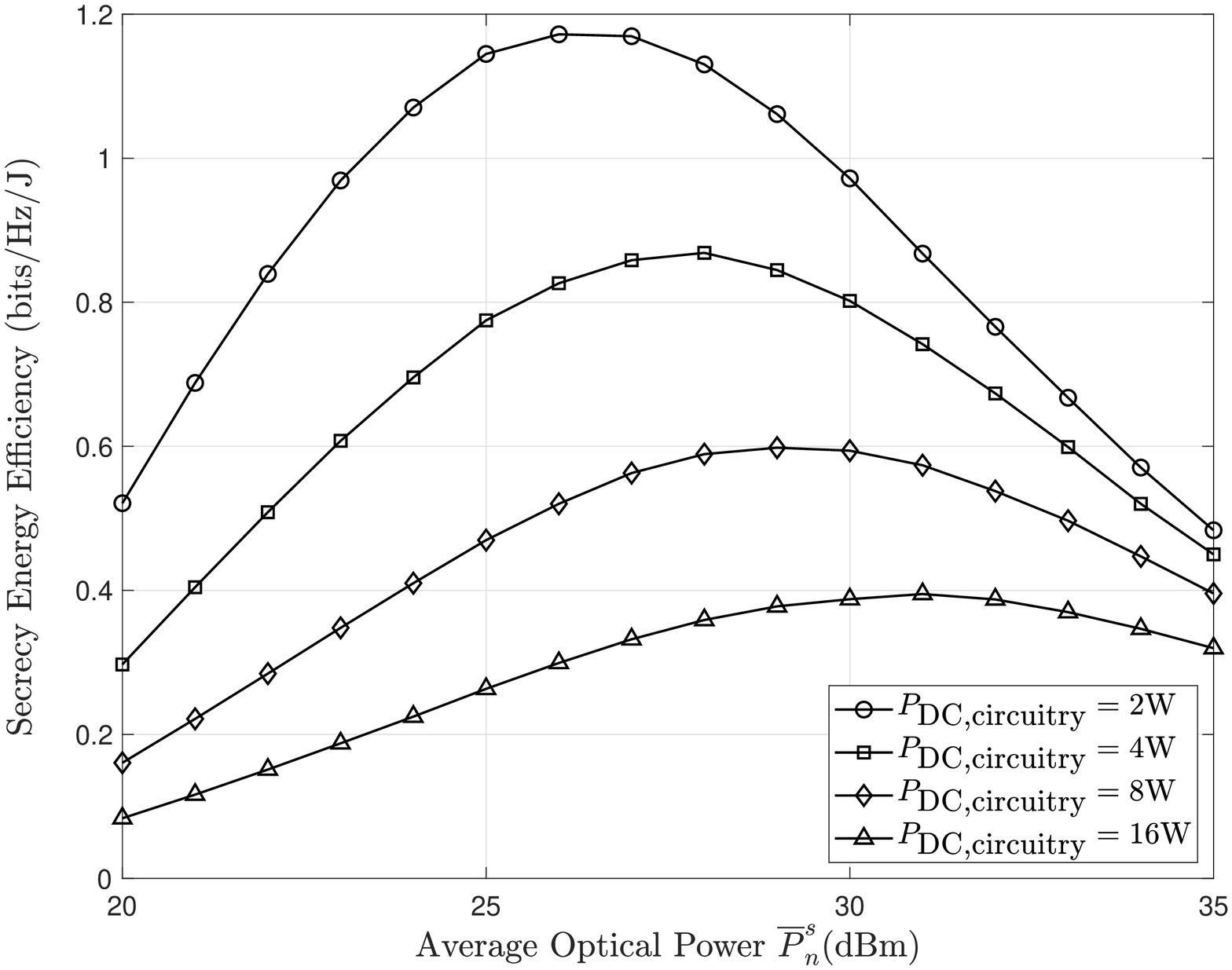}}
\caption{Energy efficiency versus  $\overline{P}^s_n$ with different  $P_{\text{DC, circuitry}}$.}
\label{EE_vs_fixed_P_VLC}
\end{figure}

In Fig. \ref{EE_vs_fixed_P_VLC}, we assess the energy efficiency with respect to each luminary's average emitted optical power $\overline{P}^s_n$ for different values of circuitry power consumption where $(N_T, K) = (4, 3)$. It is observed that the energy efficiency first increases with an increase in $\overline{P}^s_n$ until its maximum value. It then starts decreasing as $\overline{P}^s_n$ continues to increase. This phenomenon can be explained as follows. At its low value region, the increase in the achievable secrecy sum-rate due to increasing $\overline{P}^s_n$ is dominant in improving the energy efficiency. When $\overline{P}^s_n$ further increases, it becomes the dominant factor that reduces the energy efficiency as the achievable secrecy sum-rate only logarithmically increases  with $\overline{P}^s_n$. We also notice that the optimal point of $\overline{P}^s_n$ increases in accordance with $P_{\text{DC, circuitry}}$.
%there exist an optimal value of  $\overline{P}^s_n$ where the energy efficiency achieves its maximum value. In addition, this optimal value of $\overline{P}^s_n$ increases in accordance with an increase of $P_{\text{DC, circuitry}}$
%It is obvious that when the fixed DC power consumption increases, the energy efficiency of secrecy rate decreases. In addition, the increase of average optical power leads to the increase of total power consumption, and the increase of the secrecy sum rate due to the constraint (\ref{eqn:linear_constraints}) being loosen. As a result, there will be an optimum optical power level that maximizes the average energy efficiency. Our numerical result shows that the optimum optical power level increases from 26 to 31 dBm in accordance with the increase of fixed DC power consumption.

\section{Conclusions}
\label{sec:conclusion}
In this paper, we have studied an energy-efficient precoding scheme for PLS in MU-MISO VLC systems. Due to the non-convex nature of the design problem,  Dinkelbach and CCCP algorithms were employed to find a sub-optimal solution with lower complexity. Numerical results shown that by choosing ZF precoders as the initial points for the CCCP algorithm, the rate of convergence of the proposed solution could be significantly improved. It was also demonstrated that there exist an optimal value of the average emitted optical power where the energy efficiency achieves its maximum value.

\bibliographystyle{IEEEtran}
\bibliography{references}
\section*{Acknowledgement}
This work is supported by the Telecommunications Advancement Foundation (TAF) under Grant C-2020-2.
\end{document}